\documentclass[a4paper,twoside]{article}

\usepackage{graphicx}

\def\lsi{\raise0.3ex\hbox{$<$\kern-0.75em\raise-1.1ex\hbox{$\sim$}}}
\def\gsi{\raise0.3ex\hbox{$>$\kern-0.75em\raise-1.1ex\hbox{$\sim$}}}

\newcommand{\gsim}{\mathop{\gsi}}

\title{Relating Chiral Perturbation Theory and
QCD Simulations with Overlap Hypercube Fermions}

\author{W. Bietenholz$^{\rm \, a}$ and S. Shcheredin$^{\rm \, a,b}$ \\ 
 \ \ [for the $\chi$LF Collaboration] \\
\ \\
$^{\rm a}$
Institut f\"{u}r Physik, Humboldt Universit\"{a}t zu Berlin\\
Newtonstr.\ 15, D-12489 Berlin, Germany \\
\ \\
$^{\rm b}$ Fakult\"{a}t f\"{u}r Physik, Universit\"{a}t Bielefeld \\
D-33615 Bielefeld, Germany }
\begin{document}
\maketitle
\abstract{We present simulation results for lattice QCD with light pions.
For the quark fields we apply chirally symmetric lattice Dirac operators,
in particular the overlap hypercube operator, along with the standard overlap
operator for comparison. This allows us to simulate at very low pion masses.
The results are related to Random Matrix Theory and to Chiral Perturbation
Theory 
in order to extract information about the pion decay constant, 
the scalar condensate and the topological susceptibility.}

\section{Chiral Perturbation Theory}

When a continuous, global symmetry breaks spontaneously,
we obtain a continuous set of degenerate vacuum states.
Expanding around one selected vacuum, one distinguishes between excitations
to higher energy (which are identified with massive particles)
and fluctuations, which preserve the ground state energy.
The subgroups of the energy conserving symmetry group can either transfer 
the selected vacuum to a different vacuum state, or leave it simply invariant.
The number of generators relating different vacuum states corresponds 
--- according to the Goldstone Theorem --- 
to the number of massless Nambu-Goldstone
bosons (NGB) that emerge. At low energy, the NGB can be described 
by an effective theory as fields in the coset space of the
spontaneous symmetry breaking (SSB). Such effective theories still apply
if we add a small explicit symmetry breaking; we then deal with light
quasi-NGB, which dominate the low energy physics. The effective Lagrangian
${\cal L}_{\rm eff}$ contains terms of the quasi-NGB fields,
which obey the original symmetry, as well as the (explicit) symmetry 
breaking terms.
All these terms are hierarchically ordered according to some low energy
counting rules for the momenta and the quasi-NGB masses.

This concept is very general, but it was introduced in the framework of
chiral symmetry breaking in QCD. At zero quark masses QCD is assumed to 
exhibit a chiral SSB of the form
\begin{equation}
SU(N_{f})_{L} \otimes SU(N_{f})_{R} \to SU(N_{f})_{L+R} \ ,
\end{equation}
where $N_{f}$ is the number of flavours involved.
In this case, the coset space is again $SU(N_{f})$, and the corresponding
low energy effective theory is known as {\em Chiral Perturbation
Theory} ($\chi$PT) \cite{WeiGL}. 

A small quark mass supplements a slight explicit symmetry breaking,
and the quasi-NGB are then identified with the light mesons, i.e.\
the pions for $N_{f}=2$ --- and for $N_{f}=3$ also the kaons and eta
particles.

In view of our lattice study, we have to put the system into a finite volume;
we choose its shape as $V = L^{3} \times T$ $(T \geq L)$. 
Then $\chi$PT was formulated in two
regimes, with different counting rules for the terms in ${\cal L}_{\rm eff}$.
The usual case is characterised by $L m_{\pi} \gg 1$, where $m_{\pi}$ is the
pion mass, i.e.\ the lightest mass involved, which corresponds to the inverse
correlation length. This is the {\em $p$-regime}, where finite size
effects are suppressed, and one expands in the meson momenta and masses
($p$-expansion) \cite{preg}.

The opposite situation, $L m_{\pi} < 1$, is denoted as the 
{\em $\epsilon$-regime}. In that setting, an expansion in the meson momenta is
not straightforward, due to the important r\^{o}le of the zero modes.
Fortunately, the functional integral over these modes can be performed
by means of collective variables \cite{epsreg1}. There is a large gap to the
higher modes, which can then be expanded again, along with the 
meson masses ($\epsilon$-expansion) \cite{epsreg1,epsreg2}.

In both regimes, the leading order of the effective Lagrangian 
(in Euclidean space) reads
\begin{eqnarray}
{\cal L}_{\rm eff} &=& \frac{F^2}{4} \, {\rm Tr} [ \partial_{\mu} U^{\dagger}
\partial_{\mu} U ] - \frac{1}{2} \Sigma \, {\rm Tr} 
[ {\cal M} (U + U^{\dagger}) ] + \dots \ , \nonumber \\
&& U \in SU(N_{f}) \ , \ {\cal M} = {\rm diag}(m_{u}, m_{d}, (m_{s})) \ .
\end{eqnarray}
The coefficients to these terms are the 
Low Energy Constants (LEC), and we recognise $F$ and $\Sigma$ as 
the leading LEC. As it stands, $\Sigma$ and $F$ occur in the chiral limit 
${\cal M}=0$; at realistic light quark masses $F$ turns into
the pion decay constant $F_{\pi}$. Experimentally its value was
measured as $F_{\pi} \simeq 93 ~{\rm MeV}$, which is somewhat above
the chiral value of $F \simeq 86 ~{\rm MeV}$ \cite{CD}. 
$\Sigma$ is not directly accessible in experiments, 
but its value is assumed to be $\gsim (250 ~{\rm MeV})^{3}$.

The LEC are of physical importance, but they enter
the $\chi$PT as free parameters. For a theoretical prediction of their 
values one has to return to the fundamental theory, which is QCD in 
this case. There is a notorious lack of analytic tools
for QCD at low energy, hence the evaluation of the LEC is a challenge
for lattice simulations.

The LEC in nature correspond to their values at $V=\infty $, and the
$p$-regime is close to this situation. However, it is interesting that
these infinite volume values of the LEC can also be determined in
the $\epsilon$-regime, in spite of the strong finite size effects.
Actually one makes use exactly of these finite size effects
to extract the physical LEC. 
Generally, we need a long Compton wave length for the pions, $1/m_{\pi}$,
and in view of lattice simulations in the $p$-regime 
we have to use an even much larger box length $L$.
In this respect, it looks very attractive to work in the $\epsilon$-regime
instead, where we can get away with a small volume. 

However, such simulations face conceptual problems: 
first, light pions can only be realised if the lattice
fermion formulation keeps track of the chiral symmetry.
In addition, the $\epsilon$-regime has the peculiarity that 
the topology plays an important r\^{o}le \cite{LeuSmi}: 
$\chi$PT predictions for expectation values
often refer to distinct topological sectors, so it would be a drastic
loss of information to sum them up. 

\section{Lattice QCD with Chiral Fermions}

These conceptual problems could be overcome only in the recent
years. The solution is the use of a lattice Dirac operator $D$ 
which obeys the Ginsparg-Wilson relation \cite{GW}.
Its simplest form reads (in lattice units)
\begin{equation}  \label{GWR}
D \gamma_{5} + \gamma_{5} D = \frac{2}{\mu} D \gamma_{5} D \ ,
\quad \mu \gsim 1 \ ,
\end{equation}
which means that $D^{-1}$ anti-commutes with $\gamma_{5}$, up to a local
term that vanishes in the continuum limit. Even at finite lattice spacing,
this local term ($2 \gamma_{5}/\mu$ in eq.\ (\ref{GWR})) does not shift the
poles in the propagator $D^{-1}$. This fermion 
formulation has a lattice modified, but exact chiral symmetry \cite{ML}
and exact zero modes with a definite chirality. That property
provides a definition
of the topological charge by means of the fermionic index 
$\nu = n_{+} - n_{-}$ (where $n_{\pm}$ is the number of zero modes
with positive/negative chirality) \cite{Has}.

\begin{figure}[htb]
\begin{center}
\includegraphics[angle=270,scale=0.22]{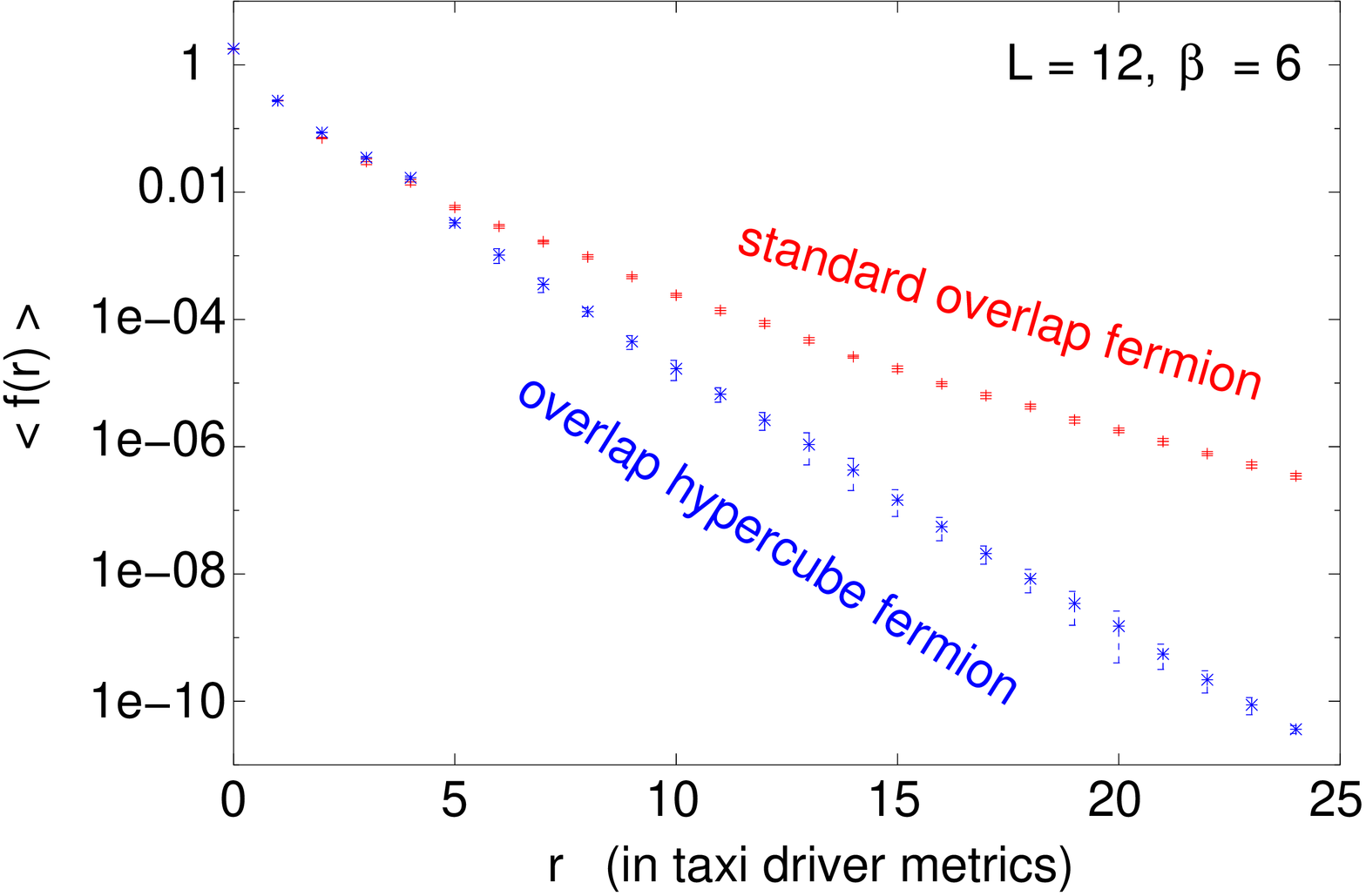}
\includegraphics[angle=270,scale=0.3]{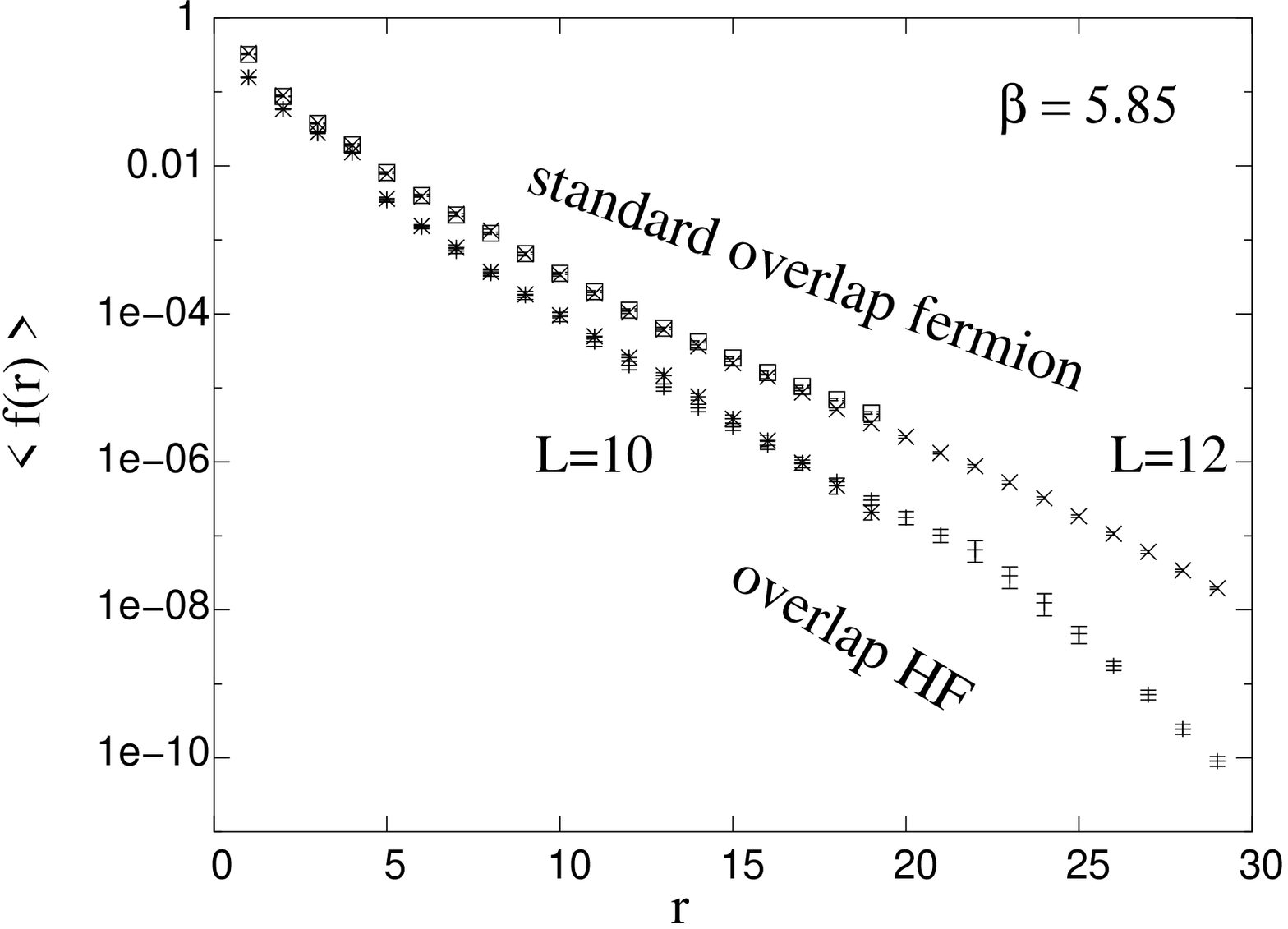}
\end{center}
\caption{\it{The function $f(r)$ measures the maximal correlation between
sink and source if they are separated by a distance $r$. 
The decay of its expectation value must be (at least)
exponential for locality to hold. Our plots show that this is
the case for both, the Neuberger fermion \cite{HJL}
and the overlap HF, at $\beta =6$ (on the left) and at $\beta =5.85$ 
(on the right) on $L^{4}$ lattices. We also see that
the decay is clearly faster for the overlap HF, i.e.\ its locality is
significantly improved.}}
\label{locfig}
\end{figure}

The simplest solution to this relation is obtained by inserting
the Wilson-Dirac operator $D_{W}$ into the so-called 
overlap formula \cite{Neu},
\begin{equation}  \label{overlap}
D_{\rm ov}^{(0)} = A_{\rm ov} + \mu \ , \quad A_{\rm ov} = \mu A_{0}
\Big( A_{0}^{\dagger} A_{0} \Big)^{-1/2} \ , \quad
A_{0} = D_{0} - \mu \ .
\end{equation}
H.\ Neuberger suggested this solution with $D_{0}=D_{W}$, and we denote
the resulting $D_{\rm ov}^{(0)}$ as the Neuberger operator (at mass zero).
We used it in all our applications presented below at $\mu = 1.6$.
It is motivated, however,
to study also the generalisation with different kernels $D_{0}$, in particular
when they already represent an approximate Ginsparg-Wilson operator 
\cite{EPJC}. We suggested to use a kernel with
couplings in a unit hypercube on the lattice (hypercube fermion, HF),
which are constructed with Grassmannian
block variable renormalisation group
transformations \cite{BBCW}. Its gauging also involves ``fat links''
\cite{HFQCD} (the same is also true for the alternative HF kernel
of Ref.\ \cite{Bern}).
Compared to $D_{W}$, multiplications with this kernel require 
more numerical work (about a factor of 15 in QCD), but part of it is gained back
immediately since the evaluation of the overlap operator --- which has to be
approximated by polynomials in practical implementations --- has
a faster convergence. There remains an overhead of about a factor 3,
but there are further gains of the resulting overlap HF
in terms of locality, rotation symmetry and scaling.
These virtues have all been tested and confirmed for free fermions
\cite{EPJC} and for the 2-flavour Schwinger model \cite{Schwing}.
In QCD we worked out such HF kernels at $\beta = 6 /g_{0}^{2} = 6$ 
\cite{HFQCD} and recently also at $\beta = 5.85$ \cite{Stani}, 
which corresponds to lattice spacings of $a \simeq 0.093~{\rm fm}$ resp.\ 
$a \simeq 0.123~{\rm fm}$ (in quenched simulations).
Again an improved locality --- see Fig.\ \ref{locfig} ---
and rotation symmetry could be confirmed, while a systematic scaling 
test is still outstanding. The axial anomaly is correctly reproduced 
in the continuum limit of any topological sector for the Neuberger
fermion \cite{DA}, as well as the overlap HF \cite{DAWB}.

A number of subtle numerical tools for simulations with overlap
fer\-mions have been elaborated \cite{algo}, but their simulation is
still computationally expensive. For the time being, only quenched
QCD simulations are possible; the inclusion of dynamical quarks \cite{dynsimu}
might be a challenge for the next generation of supercomputers.

In view of the evaluation of Low Energy Constants, the quenched 
approximation causes logarithmic finite size effects \cite{quench}.
Nevertheless it is important to explore the potential of this method,
in particular in view of the evaluation of LEC in the $\epsilon$-regime.

\section{The Pion Mass}

\begin{figure}[htb]
\begin{center}
\includegraphics[angle=270,scale=0.47]{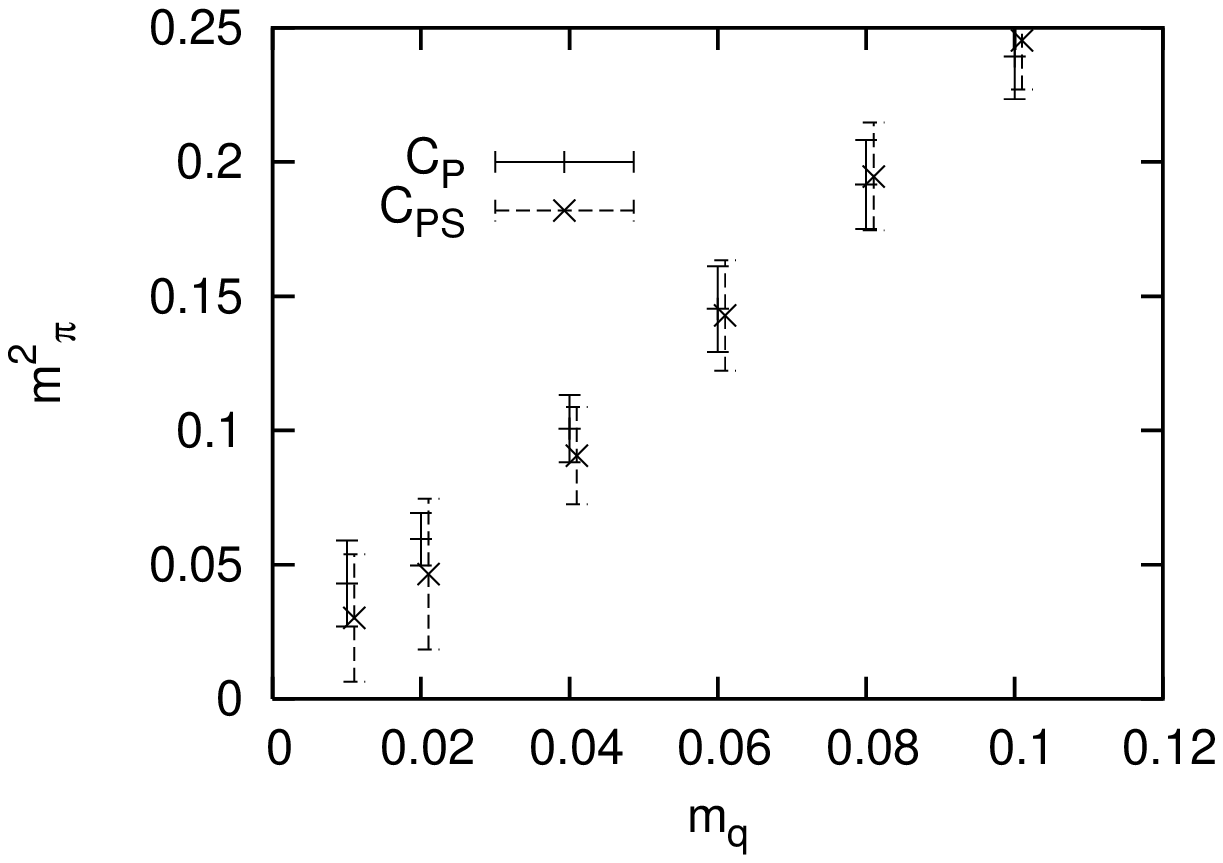}
\includegraphics[angle=270,scale=0.47]{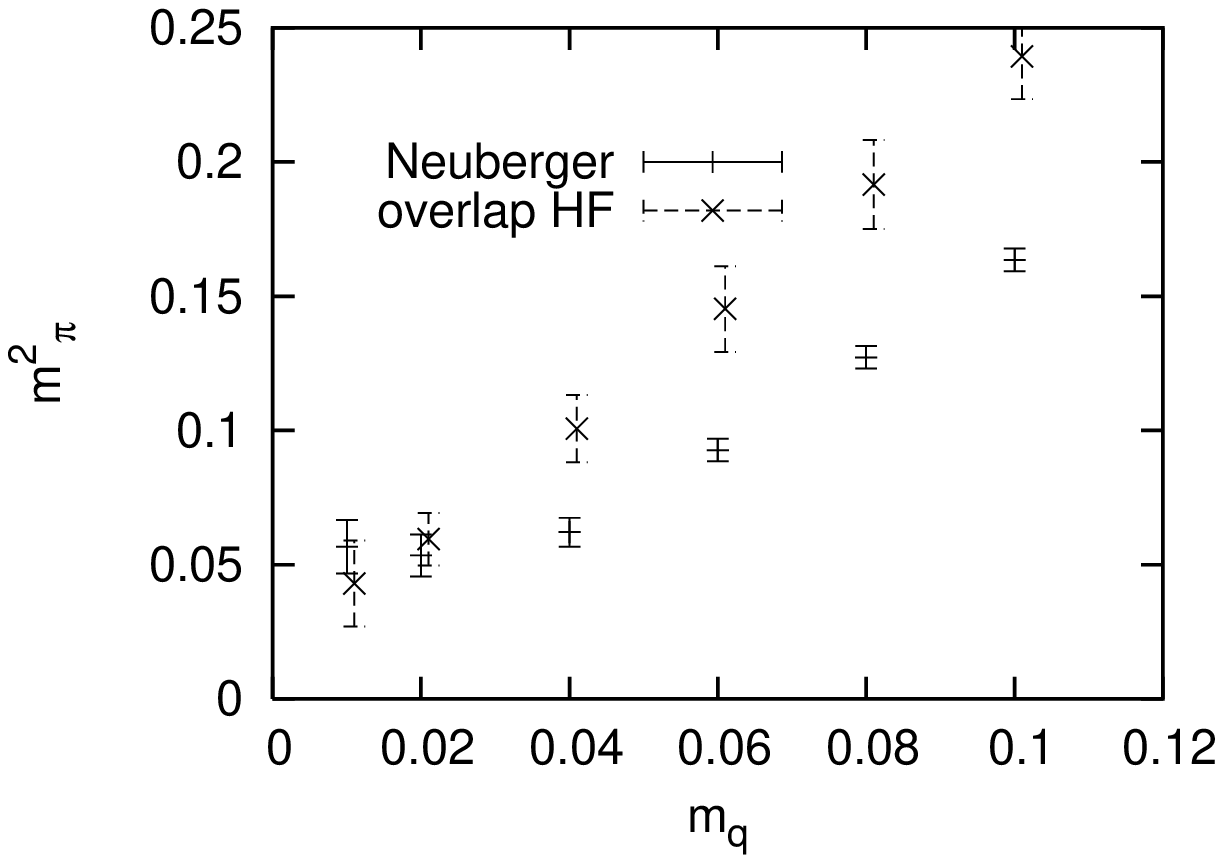}
\end{center}
\caption{\it{Left: the pion mass vs.\ the bare quark mass
(both in lattice units) for the overlap HF, evaluated from the 
temporal decay of the pseudoscalar correlator, with or without subtraction
of the scalar correlator. The lowest pion mass
in this plot (at $m_q = 0.01$)
corresponds to about $230~{\rm MeV}$.
This is a preliminary result, based on 29 propagators.
On the right we show the comparison to the Neuberger fermion,
both evaluated from $C_P$ in exactly the same way.}}
\label{pionmass}
\end{figure}

For Wilson's traditional lattice fermion formulation, it was a
insurmountable problem to reach light pion masses; due to the additive
mass renormalisation and other conceptual
problems, the pion masses always remained above about $600~{\rm MeV}$
(at least quenched).
In order to verify how close we can get to realistic pion masses 
with the overlap HF, 
we first evaluate the corresponding pion mass in the $p$-regime,
as a function of the bare quark mass $m_q$ (which we assume to be the 
same for all flavours involved). The latter is added to 
the chiral operator $D_{\rm ov}^{(0)}$ of eq.\ (\ref{overlap}) as
\begin{equation}
D_{\rm ov}(m_q ) = 
\Big( 1 - \frac{m_{q}}{2 \mu} \Big) D_{\rm ov}^{(0)} + m_{q} \ .
\end{equation}
The evaluation considers the exponential decay of the pseudoscalar correlation
function, which is deformed to a {\tt cosh} behaviour by the periodic
boundary conditions. The method can still be improved by subtracting
the scalar correlations function, i.e.\ we study the decays of both,
\begin{eqnarray}
C_{P}(t) &=& \sum_{\vec x} \langle P^{\dagger} (\vec x , t) P(0) \rangle \ , 
\qquad {\rm and} \nonumber \\
C_{PS}(t) &=& \sum_{\vec x} \langle P^{\dagger} (\vec x , t) P(0)
- S^{\dagger} (\vec x , t) S(0) \rangle \ , \label{cpcps}
\end{eqnarray}
where $P$ and $S$ are the pseudoscalar and the scalar density.
In $C_{PS}$ some contamination by topological finite size effects 
is eliminated.
The result is shown in Fig.\ \ref{pionmass} (on the left),
which illustrates that the 
data are in agreement with the expected behaviour $m_{\pi}^{2} \propto m_q$.
At $m_q = 0.01$ (in lattice units) we arrive at a
value of $m_{\pi} \approx 230 ~{\rm MeV}$, i.e.\ significantly closer
to the physical value than the simulations with Wilson fermions.
At this point, we are already
close to the transition to the $\epsilon$-regime,
$m_{\pi}L \approx 1.7$.
The corresponding pion masses for the Neuberger fermion as measured
by our collaboration with more data can be found in Ref.\ \cite{XLF}.
A measurement in exactly the same way (using the same 29 configurations)
based on $C_{P}$ is shown for comparison in Fig.\ \ref{pionmass} on the right.

\section{The Topological Susceptibility}

\begin{figure}[htb]
\begin{center}
\includegraphics[angle=270,scale=0.33]{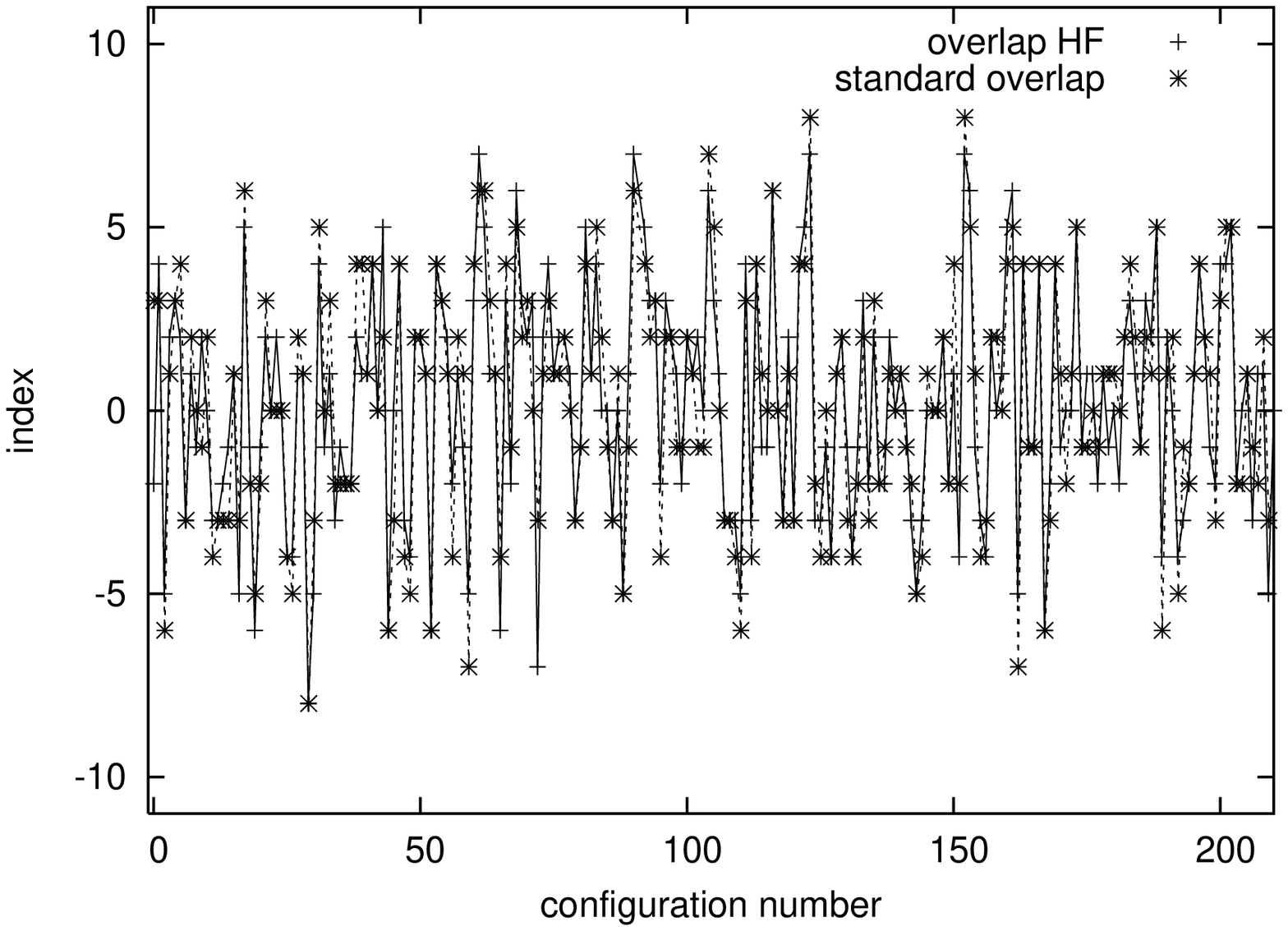}
\includegraphics[angle=270,scale=0.4]{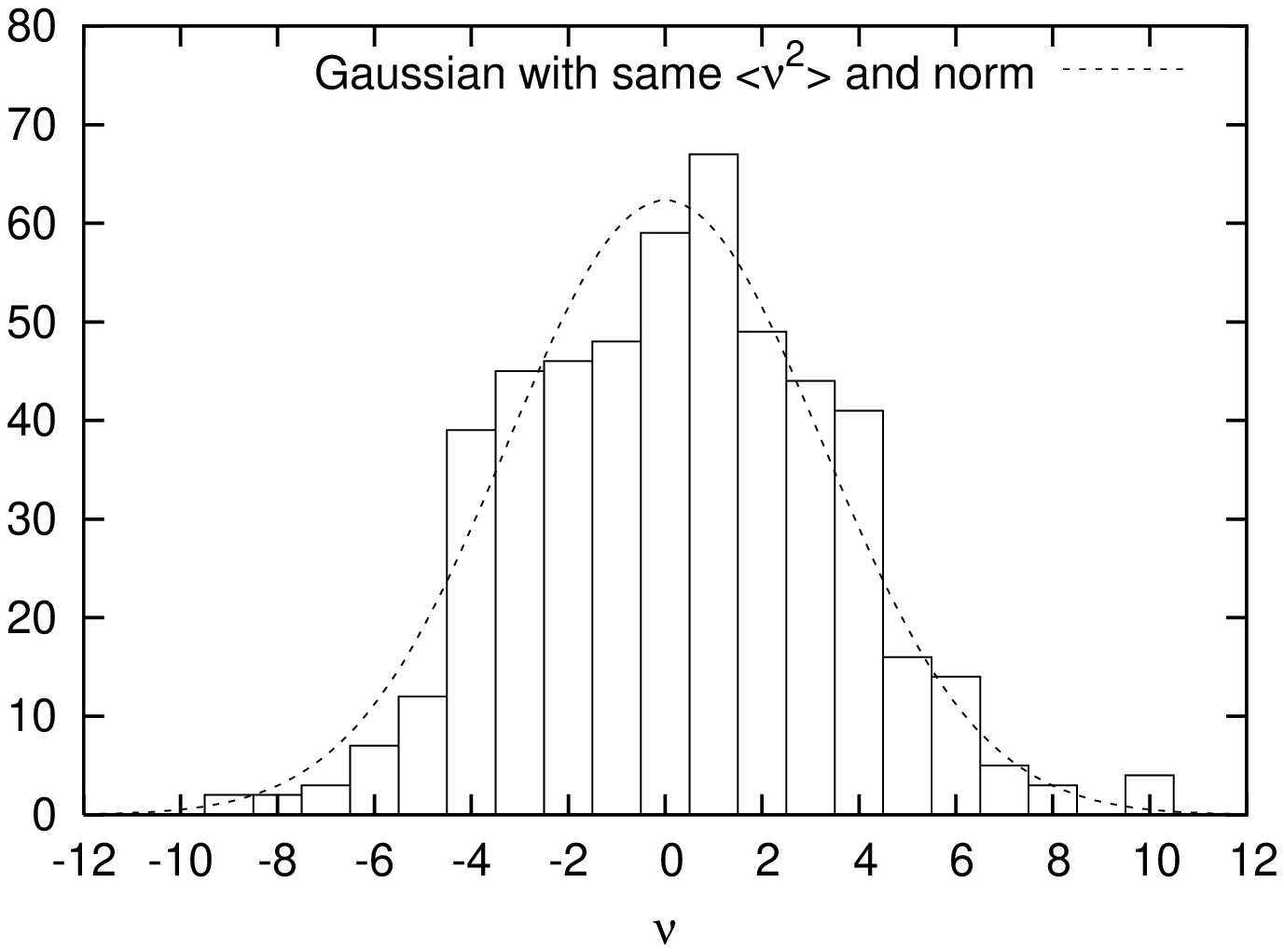}
\end{center}
\caption{\it{On the left we show part of the index histories
for the overlap HF and the Neuberger fermion, for the same configurations
on a $12^3 \times 24$ lattice at $\beta = 5.85$. The plot on the right shows
the distribution of topological charges on a $16^3 \times 32$
lattice at $\beta =6$, measured with the index of the Neuberger operator
at $\mu = 1.6$ on 506 configurations. We see a decent agreement 
with a Gaussian distribution. Its width represents the topological 
susceptibility, see Fig.\ \ref{susfig}.}}
\label{indexfig}
\end{figure}

In the quenched approximation, the topological charge (which is identified
with the fermion index $\nu$) 
does not depend on the quark mass. Its statistical distribution
is expected to be Gaussian. Fig.\ \ref{indexfig} 
(on the left) shows part of our index history for the overlap HF 
and the Neuberger fermion. They deviate a little,
$ \langle | \nu_{\rm ov-HF} - \nu_{\rm N} | \rangle \simeq 0.89(4)$.
The plot on the right shows, as an example,
our index histogram measured with the Neuberger operator at $\mu =1.6$,
which is reasonably consistent with a Gaussian.

From the point of view of the naive, constituent quark model,
the $\eta '$ meson is amazingly heavy 
($m_{\eta '} \simeq 958~{\rm MeV}$).
An explanation for this
property was given based on topological windings of the Yang-Mills
gauge field. In particular, the Witten-Veneziano formula
relates $m_{\eta '}$ to the topological susceptibility $\chi_{t}$
in the leading order of a $1/N_c$ resp.\ $N_f/N_c$ expansion 
($N_c$ is the number of colours),
\begin{equation}
m_{\eta '}^{2} = \frac{2 N_{f}}{F_{\pi}^{2}} \chi_{t} \ , \quad
\chi_{t} = \frac{1}{V} \langle \nu^{2} \rangle \ ,
\quad \nu : {\rm topological~charge}.
\end{equation}
In this formula $\chi_{t}$ is understood as a property of pure gauge theory
(hence it is sensible to define it with the quenched fermion index),
whereas the other terms refer to full QCD. A recent large-scale study 
with Neuberger fermions \cite{toposus} arrived in the continuum extrapolation 
at $\chi_{t} = (191 \pm 5 ~{\rm MeV})^4$,
or --- in dimensionless units ---  $\chi_{t} r_{0}^{4} = 0.059(3)$ (where
$r_{0}$ is the Sommer scale, which translates lattice quantities into
physical units). This is consistent with a heavy $\eta '$ meson.

\begin{figure}[htb]
\begin{center}
\includegraphics[angle=270,scale=0.39]{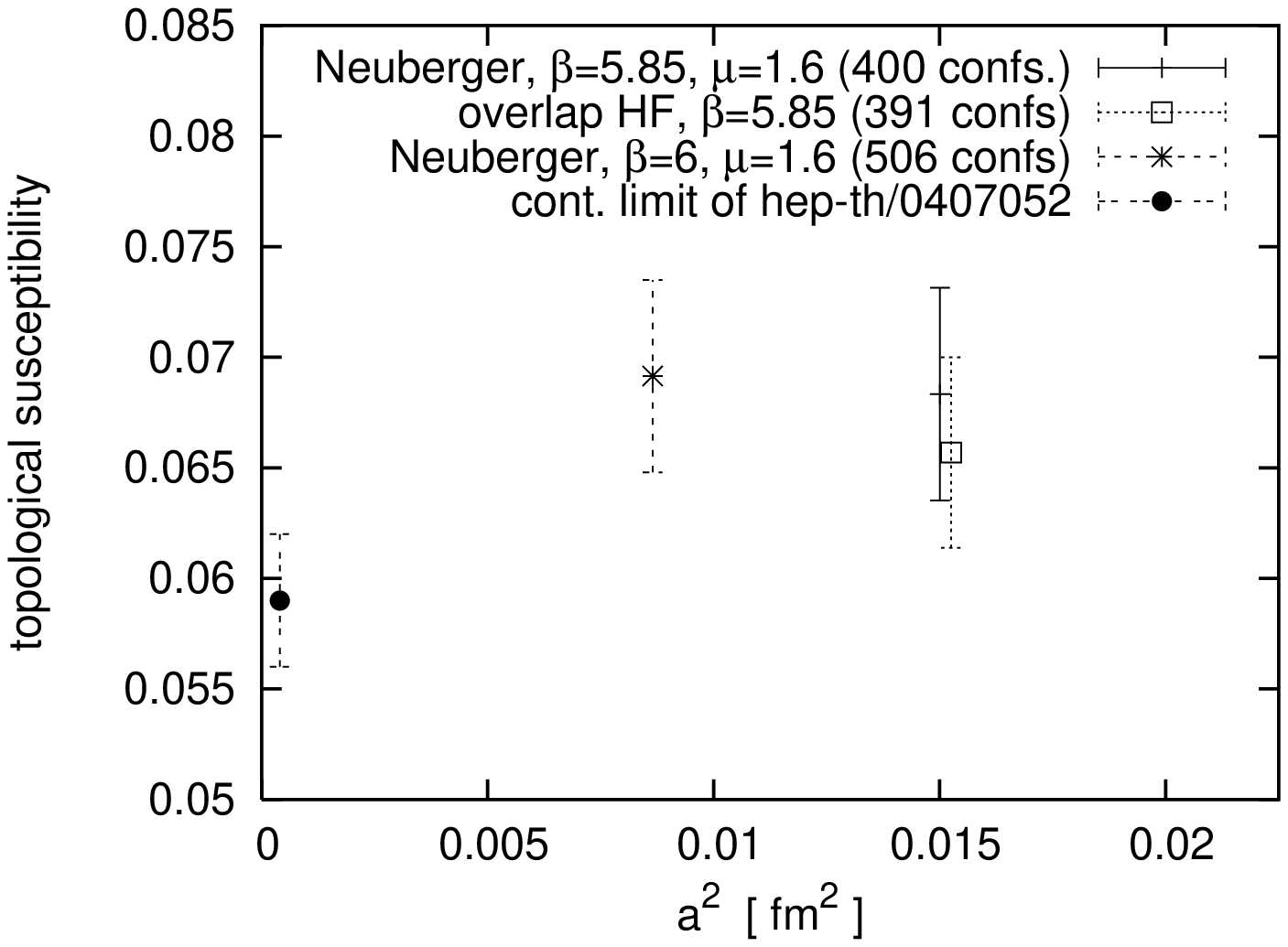}
\includegraphics[angle=270,scale=0.39]{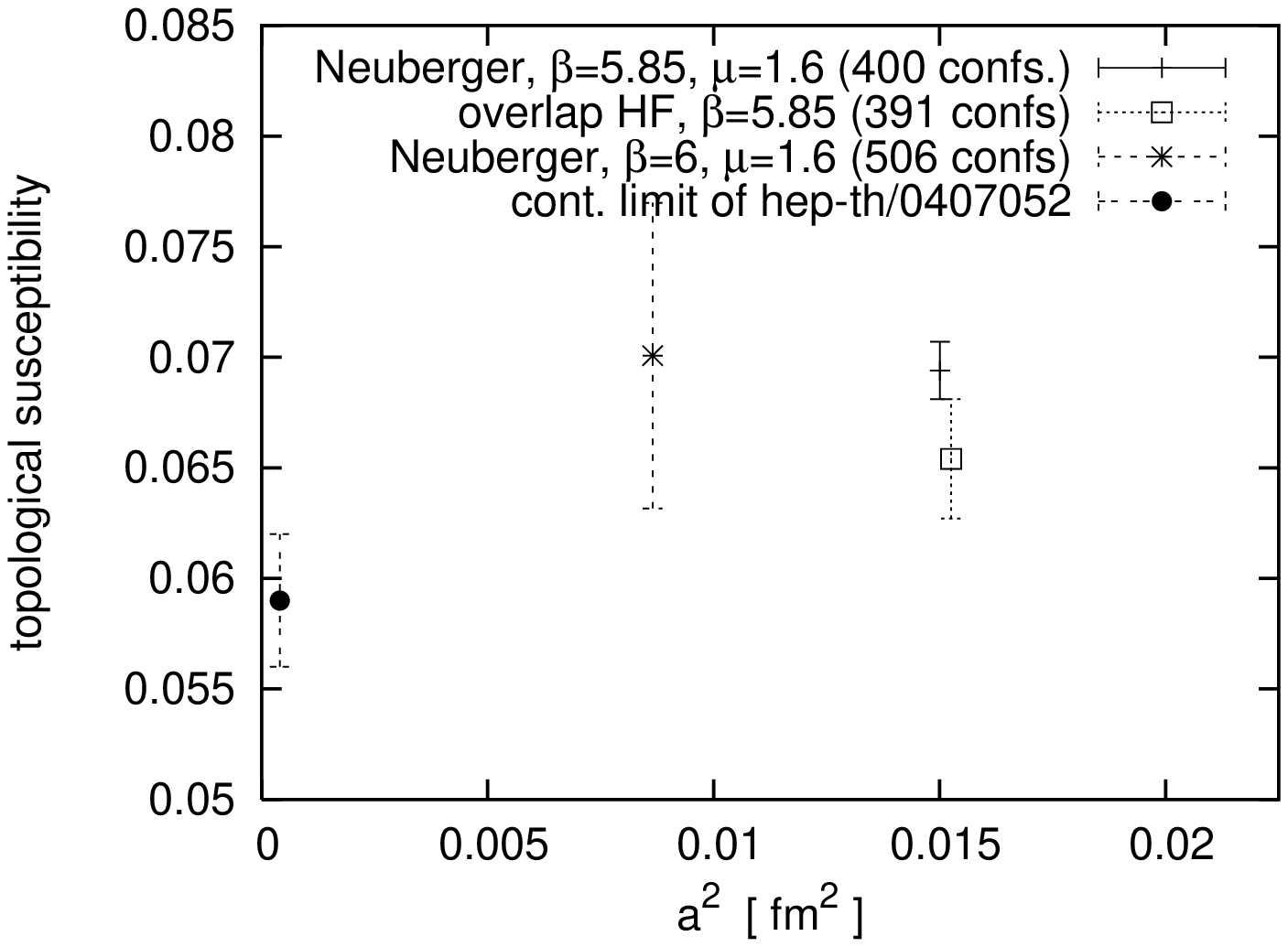}
\end{center}
\caption{\it{The topological susceptibility $\chi_{t} r_{0}^{4}$
at different lattice spacings
in the same volume $V = (1.48~{\rm fm})^{3} \times 2.96 ~ {\rm fm}$.
On the left we show our result with the standard statistical evaluation. 
Alternatively we cut the measured distributions at
$| \nu | = 1.5$, $2.5, \dots ,5.5$ and matched them to a Gaussian each time,
so that in each case a larger statistics is involved. The results obtained
in that way is shown on the right. So far our value for $\chi_{t}$
is a little higher than the one of Ref.\ \cite{toposus}, which is,
however, based on a larger statistics.}}
\label{susfig}
\end{figure}

In Fig.\ \ref{susfig} we show our preliminary result for
$\chi_{t} r_{0}^{4}$ in a physical volume of $(1.48~{\rm fm})^{3} \times
2.96 ~ {\rm fm}$ at lattice spacings $a=0.123~{\rm fm}$ and $a=0.093~{\rm fm}$.
We include results for the Neuberger operator, as well as the overlap
HF, and we also show the continuum extrapolation
of Ref.\ \cite{toposus} for comparison.
Our present data suggest a trend to a somewhat larger value of
$\chi_{t} r_{0}^{4}$, though the overlap HF is closer to the result
of the literature. More statistics will be required to clarify if we arrive
at a result consistent with Ref.\ \cite{toposus}.

\section{Determination of $\Sigma$}

Chiral Random Matrix Theory (RMT) introduces a pseudo Dirac operator
\begin{equation}
D_{\rm RMT} = \left( \begin{array}{cc}
0 & i W \\ i W^{\dagger} & 0 \end{array} \right) \ ,
\quad W : {\rm complex ~} n_{+} \times n_{-} {\rm ~ random~matrix} .
\end{equation}
In this way, QCD is simplified to a Gaussian distribution of 
fermion matrix elements. The hope is to capture nevertheless
some properties of QCD in the $\epsilon$-regime. In fact, the 
Leutwyler-Smilga spectral sum rules \cite{LeuSmi} were successfully 
reproduced in this way \cite{VZ}.

There are also explicit conjectures for the probability distribution
$\rho$ of the low lying eigenvalues $i\lambda$ of the continuum
Dirac operator. One introduces the dimensionless variable
$z = \lambda \Sigma V$ and considers the spectral density
\begin{equation}  \label{specdense}
\rho_{s}(z) = \left. \frac{1}{\Sigma V} \rho \Big( \frac{z}{\Sigma V} \Big)
\right\vert_{V = \infty} = \sum_{\nu = -\infty}^{\infty} 
\rho_{s}^{(\nu )} (z) \ .
\end{equation}
Formally this term is taken at $V = \infty$ (so one deals with a 
continuous spectrum), although the prediction refers to the 
$\epsilon$-regime. Note, however, that the quark mass is set to
zero. This shows that the conjecture involves a number of 
assumptions, and a test against lattice data is motivated. The last step
in eq.\ (\ref{specdense}) is a decomposition into the contributions
of the topological sectors, which only depend on $| \nu |$.
In a fixed sector we can further consider the individual densities of
the leading non-zero eigenvalues,
\begin{equation}
\rho_{s}^{(\nu )} (z) = \sum_{n = 1,2,3,\dots}
\rho_{n}^{(\nu )} (z) \ .
\end{equation}
Chiral RMT provides explicit predictions for the leading densities
$\rho_{n}^{(\nu )} (z)$ \cite{lowEV}, which are supposed to hold 
up to some energy (depending on the volume). These predictions have
been compared to lattice data for overlap fermions
in Refs.\ \cite{EV1}, which did agree
for not too small volumes, up to some value of $z$. 
\footnote{The work by Giusti et al.\ is the most extensive study
in this context.}
The only free
parameter involved in these fits is the scalar condensate $\Sigma$.
The successful fits provide therefore also a value
for $\Sigma$, which is found in the range where it is generally
expected. At least $\rho_{1}^{(\nu )} (z)$ agreed well with the RMT
prediction at $|\nu | = 0,1,2$ for $L \gsim 1.1~{\rm fm}$.

\begin{figure}[htb]
\begin{center}
\includegraphics[angle=270,scale=0.4]{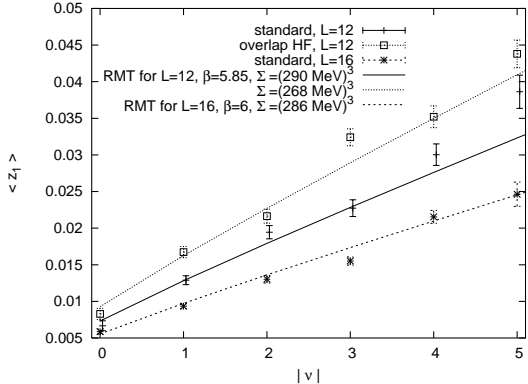}
\end{center}
\caption{\it{We show our data for the leading non-zero Dirac eigenvalue
$\langle \lambda_{1} \rangle$ (resp.\ $\langle z_{1} \rangle
= \Sigma V \langle \lambda_{1} \rangle$) in 
the topological sectors $|\nu | = 0, 1, \dots ,5$, in a volume
$(1.48~{\rm fm})^{3} \times 2.96 ~ {\rm fm}$. For the Neuberger operator
we measured at lattice spacings $a=0.093~{\rm fm}$ and $a=0.123~{\rm fm}$,
and in the latter case we also used the overlap HF operator. The optimal fit
leads to a value for $\Sigma$ in each case, which is indicated in the
plot.}}
\label{sigmafig}
\end{figure}

In Fig.\ \ref{sigmafig} we show a new result, for
the mean value of the leading non-zero eigenvalue, $\langle \lambda_{1}
\rangle$, 
\footnote{The spectrum of our Ginsparg-Wilson operators is actually located
on a circle in the complex plane with radius and centre $\mu$. 
For a comparison with the RMT predictions in the continuum, we map 
this circle 
with a M\"{o}bius transform
onto the imaginary axis.}
in the topological sectors $|\nu | = 0, 1, \dots , 5$
on different lattices with different overlap operators, but again in
a fixed physical volume. We then fit the RMT predictions for the 
optimal value of $\Sigma$ to these data, which yields values
in the range $\Sigma = (268~{\rm MeV})^{3} \dots (290~{\rm MeV})^{3}$.

At this point we mention that also staggered fermions have a remnant
chiral symmetry on the lattice, though not with the full number of generators
(in contrast to Ginsparg-Wilson fermions). Thus they are also free of additive
mass renormalisation. The standard formulation turned out to be ``topology
blind'', but a sensitivity to $| \nu |$ --- similar as the one shown in Fig.\
\ref{sigmafig} --- can be obtained by a suppression of the mixing between
its pseudo-flavours \cite{stagg}.

\section{Mesonic Correlation Functions}

To obtain direct predictions for quenched simulations results
with chiral lattice fermions, also quenched $\chi$PT has been
worked out for mesonic correlation functions \cite{qXPT}.
The vector correlator vanishes to all orders, and the scalar and
pseudoscalar correlators involve additional LEC in the leading order, 
which are due to quenching. This is not the case for the axial vector
correlator, where a parabolic (rather than a {\tt cosh}) behaviour
was predicted in the $\epsilon$-regime. In fact, this shape could
be observed \cite{AA} if the spatial box length exceeds about
$L \gsim 1.1~{\rm fm}$, 
i.e.\ the same bound that we encountered in Section 5.
This also allowed for a determination of 
$F_{\pi}$. On the other hand, the sensitivity of the curve to 
$\Sigma$ is too weak to evaluate it in this way.
We further noticed that the Monte Carlo histories
in the sector $\nu =0$ are plagued by strong spikes, 
which are related to the significant density of very small Dirac eigenvalues
\cite{AA} (at higher charges, this density is suppressed).
Hence a brute force measurement in the topologically 
neutral sector would require a tremendous statistics
(see also Ref.\ \cite{HJLl}). A strategy to avoid this problem
was proposed and applied in Ref.\ \cite{LMA}. 
Work is also in progress to
establish a lattice gauge action, which preserves the topological charge
over long periods in the Monte Carlo history \cite{topogauge}. 
This would help us to measure expectation values in a specific sector ---
which is desirable in the $\epsilon$-regime. With the standard plaquette
gauge action (that we are using so far) it is tedious to collect
statistics in a specific topology.

\begin{figure}[htb]
\begin{center}
\includegraphics[angle=270,scale=0.325]{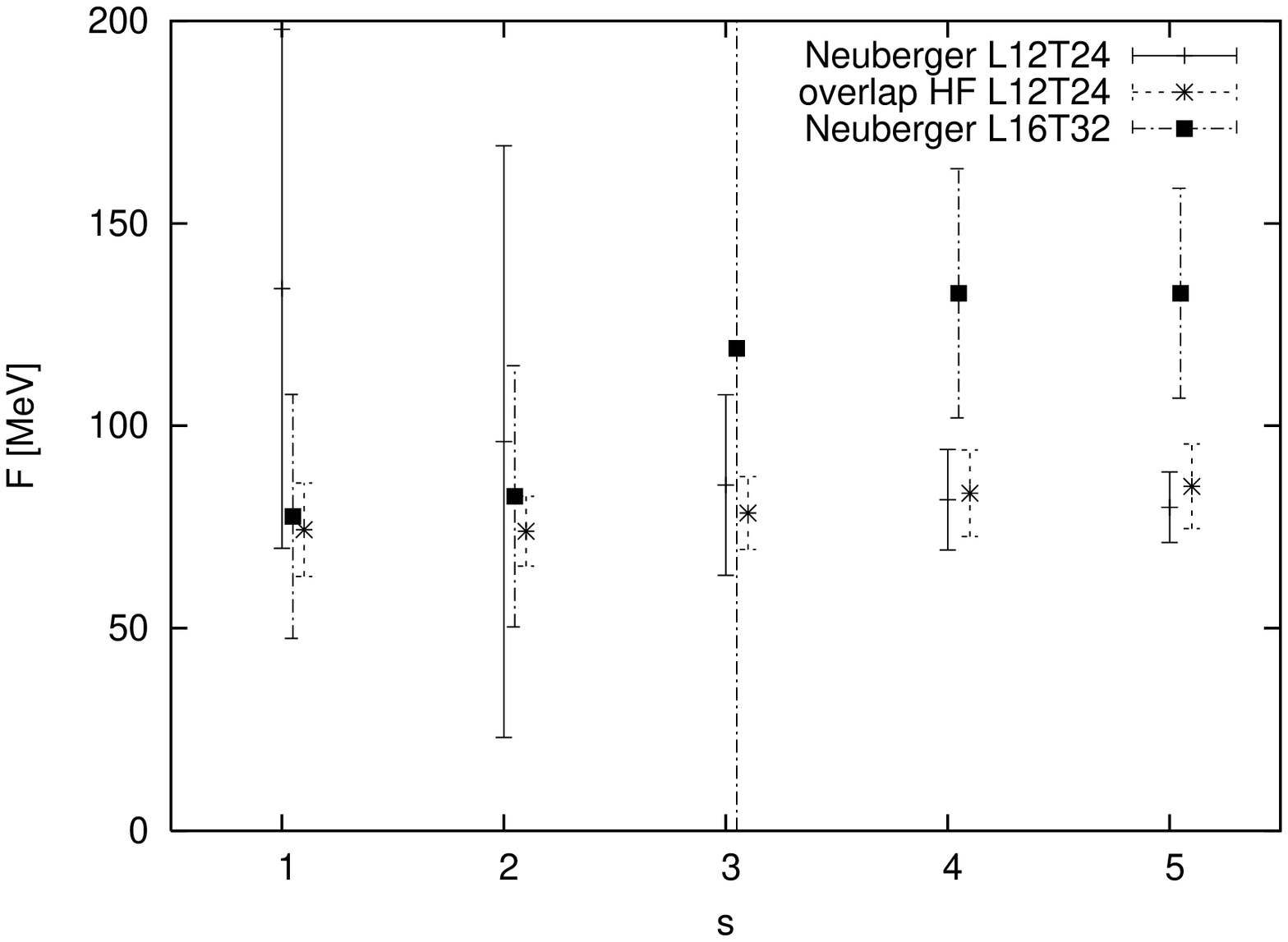}
\includegraphics[angle=270,scale=0.325]{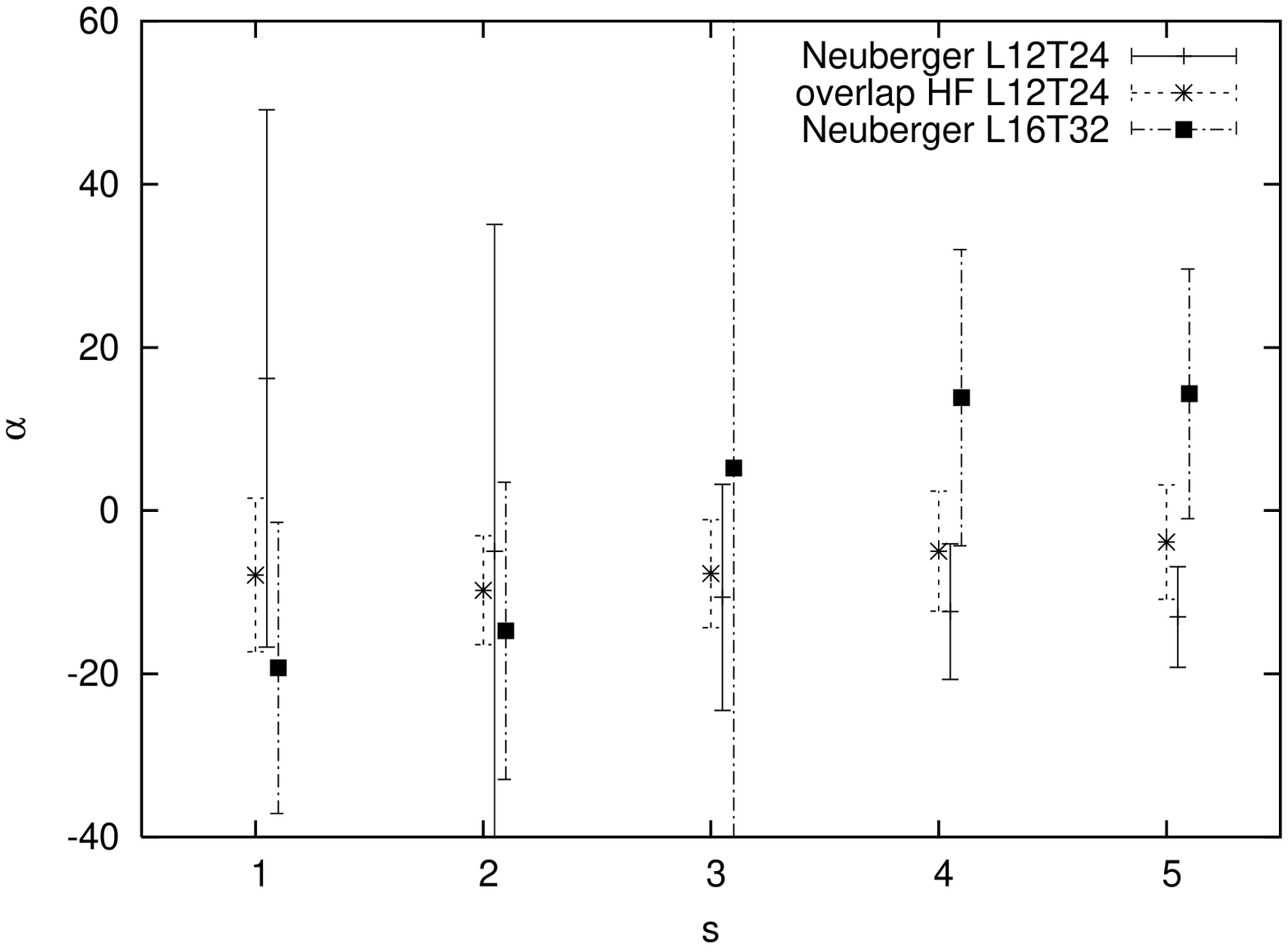}
\end{center}
\caption{\it{Results for $F$ and $\alpha$
(LEC in the leading order of quenched $\chi$PT),
based on our data for the zero-mode contribution to the pseudo-scalar
correlation function in the topological sectors $|\nu |=1$ and $2$.
In both cases, we obtain a stable behaviour and a decent extrapolation 
down to a small fitting range $s$ only for the overlap HF.
The statistics for these plots is given in Table \ref{stati0}.}}
\label{Fpifig}
\end{figure}

Here we turn our attention to another procedure, which only
considers the {\em zero-mode contributions} to the correlators.
Following Ref.\ \cite{zeromodes} we focus on the pseudoscalar correlator,
where some re-definitions allow us to study only $F$ and one
quenching specific LEC called $\alpha$ in the leading order
of quenched $\chi$PT. We distinguish the connected and the disconnected
zero-mode contributions,
\begin{eqnarray}
&& \hspace*{-13mm}
{\cal C}_{|\nu |} (x) = \langle 
v_{j}^{\dagger}(x) v_{i}(x)
v_{i}^{\dagger}(0) v_{j}(0) \rangle \, , \
{\cal D}_{|\nu |} (x) = \langle 
v_{i}^{\dagger}(x) v_{i}(x)
v_{j}^{\dagger}(0) v_{j}(0) \rangle \, ,
\end{eqnarray}
where 
$i,j$ are summed over all the zero-modes,
$D_{\rm ov}^{(0)} v_{i}=0$. 
It is difficult to fit these quantities directly,
but it is easier to fit the data to the leading order in the
expansion around the minimum in the time coordinate $t$ \cite{zeromodes}
(after summation over the spatial lattice sites $\vec x$).
This minimum is at $T/2$, hence we expand in $s = t - T/2$,
\begin{equation}  \label{zeroapprox}
\frac{1}{L^{2}} \frac{d}{ds} {\cal C}_{| \nu |} (s) \simeq \frac{s}{T} \cdot 
\tilde {\cal C}_{| \nu |} \ , \quad
\frac{1}{L^{2}} \frac{d}{ds} {\cal D}_{| \nu |} (s) \simeq \frac{s}{T} \cdot
\tilde {\cal D}_{| \nu |} \ ,
\end{equation}
where we neglect ${\cal O}((s/T)^{3})$. The combined fit of our data
in the sectors $|\nu |=1,2$ to $\tilde {\cal C}$ and $\tilde {\cal D}$ 
--- at different values of $s$ --- 
leads to the results for $F$ and $\alpha$ shown 
in Fig.\ \ref{Fpifig} \cite{Stani}, see also Ref.\ \cite{KIN}.
(In these fits we also made use of our measured
results for $\langle \nu^{2} \rangle$.) 
Since eqs.\ (\ref{zeroapprox}) only hold up to 
${\cal O}((s/T)^{3})$ we should extrapolate down to small $s$.
We only obtained a stable extrapolation, with acceptable errors,
for the overlap HF, but not for the Neuberger operator on two different
lattice spacings --- although the statistics is similar, 
see Table \ref{stati0}.
\begin{table}
\begin{center}
\begin{tabular}{|c|c|c|c|c|c|}
\hline
Dirac operator & $\beta$ & lattice size && $|\nu |=1$ &
$|\nu |=2$ \\
\hline
Neuberger & 5.85 & $12^{3}\times 24$ && ~95 & 79 \\
\hline
overlap HF & 5.85 & $12^{3}\times 24$ && ~94 & 83 \\
\hline
Neuberger & ~6 & $16^{3}\times 32$ && 115 & 95 \\
\hline
\end{tabular}
\end{center}
\caption{\it{The statistics of configurations at $|\nu |=1$ and $2$,
for different lattices and overlap operators, which was included in our
study of the zero-mode contributions to the pseudoscalar correlator.
The result is shown in Fig.\ \ref{Fpifig}.}}
\label{stati0}
\end{table}

The result of these extrapolations agrees within the 
(sizable) errors with Ref.\ \cite{zeromodes}.
Note that these are values for the bare parameters.
In particular the renormalisation of $F$
involves the factor $Z_{A}$, which amounts to about
$1.45$ in the chiral limit for our parameters of the 
Neuberger operator at $\beta = 5.85$ \cite{XLF}.

\section{Conclusions}

We discussed the potential of relating simulation results for lattice QCD
with chiral fermions to the predictions by $\chi$PT. 
For the first time we applied the overlap HF in this context
and compared its results also with Neuberger's standard overlap
operator. We gave preliminary results for the pion mass in the
$p$-regime. In the $\epsilon$-regime we are interested in 
the determination of the LEC that appear in the effective Lagrangian.

This project is still on-going, as part of the efforts of the $\chi$LF
collaboration to simulate QCD close to the chiral limit. For the approach
discussed here we saw that we can go down close to $200~{\rm MeV}$
with the pion mass. The evaluation of the LEC in the leading order seems
feasible, though a larger statistics is required for this purpose.
So far we can report that data obtained in the $\epsilon$-regime
--- i.e.\ in small volumes --- can indeed be matched
with the analytical predictions by chiral RMT and quenched $\chi$PT. 
We expect the same property also for dynamical chiral quarks, 
once our machines are powerful enough to simulate them.\\

\vspace*{-2mm}
\noindent
{\it
{\bf Acknowledgement} \ 
We are indebted to the whole
$\chi$LF collaboration for its cooperation, in particular to Mauro
Papinutto and Carsten Urbach for providing us with highly optimised 
eigenvalue routines, and to Kei-ichi Nagai for contributing
to the zero-mode study in an early stage. 
W.B.\ thanks the organisers of the 
Titeica-Markov Symposium for their kind hospitality.
This work was supported by the Deutsche Forschungsgemeinschaft 
through SFB/TR9-03. 
The computations were performed on the IBM p690 
clusters of the ``Norddeutscher Verbund f\"ur Hoch- und 
H\"ochstleistungsrechnen'' (HLRN) and at NIC, 
Forschungszentrum J\"{u}lich.
Finally we thank Hinnerk St\"{u}ben for 
his advice on the parallelisation of our codes.
}

\end{document}